# Nematic twist-bend phase with nanoscale modulation of molecular orientation

V. Borshch[1], Y.-K. Kim[1], J. Xiang[1], M. Gao[1], A. Jákli[1], V. P. Panov[2], J. K. Vij[2],

C. T. Imrie[3], M. G. Tamba[4], G. H. Mehl[4], O.D. Lavrentovich[1*]

**Affiliations:**

[1] Liquid Crystal Institute and Chemical Physics Interdisciplinary Program, Kent State University, Kent, OH 44242, USA

[2] Department of Electronic and Electrical Engineering, Trinity College, University of Dublin, Dublin 2, Ireland

[3] Department of Chemistry, School of Natural and Computing Sciences, University of Aberdeen AB24 3UE, Scotland, UK

[4] Department of Chemistry, University of Hull, Hull, HU6 7RX, UK

*olavrent@kent.edu

## ABSTRACT

A state of matter in which molecules show a long-range orientational order and no positional order is called a nematic liquid crystal. The best known and most widely used (for example, in modern displays) is the uniaxial nematic, with the rod-like molecules aligned along a single axis, called the director. When the molecules are chiral, the director twists in space, drawing a right-angle helicoid and remaining perpendicular to the helix axis; the structure is called a chiral nematic. In this work, using transmission electron and optical microscopy, we experimentally demonstrate a new nematic order, formed by achiral molecules, in which the director follows an oblique helicoid, maintaining a constant oblique angle with the helix axis and experiencing twist and bend. The oblique helicoids have a nanoscale pitch. The new twist-bend nematic represents a structural link between the uniaxial nematic (no tilt) and a chiral nematic (helicoids with right-angle tilt).

INTRODUCTION

Nematic liquid crystals with fluid-like arrangements of molecules that pack parallel to each other are widely used in display and other applications because of the unique combination of orientational order and fluidity. In the uniaxial nematic (N) phase, rod-like molecules are on average parallel to the single director $\hat{\mathbf{n}}$, but their centers of mass are arranged randomly, as in an isotropic fluid, Fig.1a. The director is a non-polar entity, $\hat{\mathbf{n}} \equiv -\hat{\mathbf{n}}$, even if the molecules have dipole moments. Chiral molecules prefer to twist with respect to each other, forcing $\hat{\mathbf{n}}$ to follow a right-angle helicoid, either left-handed, or right-handed, Fig.1c. In 1973, R.B. Meyer[1] predicted that polar molecular interactions that favor bend deformations might lead to a twist-bend nematic ($N_{tb}$) phase, in which the director draws an oblique helicoid, maintaining a constant oblique angle $0 < \theta_0 < \pi/2$ with the helix axis $z$:

$$\hat{\mathbf{n}} = \left( \sin\theta_0 \cos\varphi, \sin\theta_0 \sin\varphi, \cos\theta_0 \right); \qquad (1)$$

here $\varphi = t_{tb} z$ is the azimuthal angle, $t_{tb} = 2\pi / p_{tb}$, $p_{tb}$ is the pitch of the helicoid, Fig.1b. Note that Eq. (1) describes also N (when $\theta_0 = 0$) and $N^*$ (when $\theta_0 = \pi/2$) phases, Fig.1. Unlike the case of $N^*$, formation of $N_{tb}$ does not require molecular chirality, thus one should expect it to contain coexisting domains of left and right chirality[2]. Instead of chirality, $N_{tb}$ can be facilitated by bent (banana-like) shapes of molecules, as was demonstrated analytically by Dozov[2] and Shamid et al[3], and in molecular simulations by Memmer[4]. A similar structure, but with the hexatic order coupled to twist-bend deformation, has been predicted by Kamien[5].

Experimentally, no $N_{tb}$ phase was reported for the bent-core materials[6]. Instead, some unusual behavior, including a first-order phase transition between two seemingly uniaxial nematic phases, was detected in materials formed by polymer[7,8] and dimer molecules[9-20] in which rigid cores are connected by a flexible aliphatic tail with an odd number of methylene groups. While the high-temperature phase was easily identifiable as a standard uniaxial N phase, the nature of the low-temperature phase (often denoted $N_x$) remains a subject of intensive exploration, revealing hints that are consistent with the $N_{tb}$ structure. For example, X-ray diffraction (XRD) shows no periodic variation of the electronic density in the low-temperature phase[11,14], suggesting that the molecular centers of mass are distributed randomly in space; this excludes the smectic type of order. On the other hand, the optical textures show features such as focal conic domains[14,20]. As demonstrated by Friedel in 1922[21], focal conics appear in liquid crystals with one-dimensional positional order. This order can be caused by periodically changing density, as in smectics, or by "wave surfaces" of the director twist, as in N*, with no density modulation[22]. Focal conics should be expected[14] in $N_{tb}$, since $p_{tb}$ is fixed by the molecular interactions that favor twist-bend packing. Unlike the case of N*, in which the large pitch makes it possible to trace the helicoidal packing optically, no such clear evidence was presented so far for the $N_{tb}$ candidates. Macroscopic stripes with a period in the range of (1-100) μm often observed in dimer materials[10,11,14] do not represent a thermodynamically stable state, as the period depends strongly on the cell thickness[11]. Recently, N. Clark's group established by FFTEM a periodic director modulation in the $N_{tb}$ phase of cyanobiphenyl material M1 (Fig.2a)[19]; the period of 8 nm was not associated with the smectic periodicity. Another important feature, a structural chirality of M1 at the short time scales of NMR response, was demonstrated by L. Beguin et al[15].

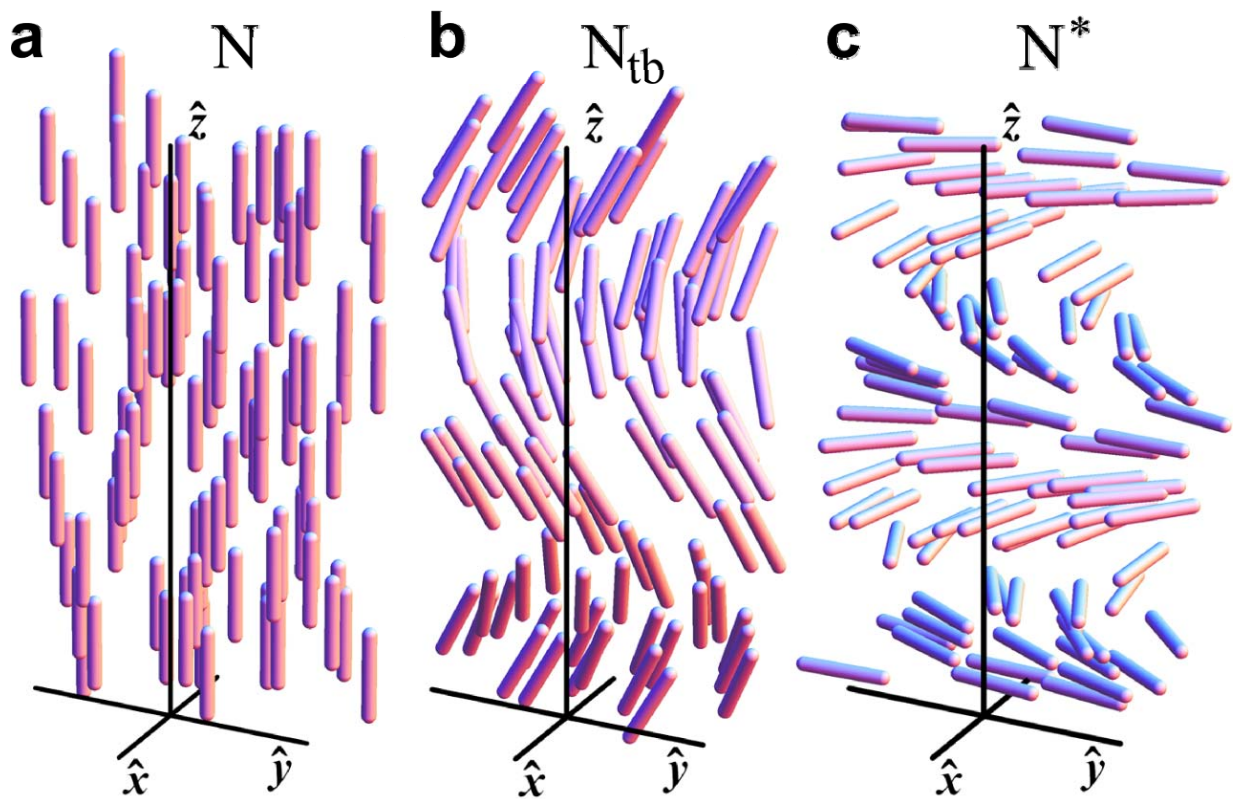

**Figure 1 | Schematics of local director arrangements in nematics**. (a) Nematic N phase, uniaxial alignment, $\theta_0 = 0$, (b) Twist-bend $N_{tb}$, with oblique helicoid $0 < \theta_0 < \pi/2$, and (c) Chiral nematic $N^*$ phase, right helicoid, $\theta_0 = \pi/2$ (the twist is right-handed or left-handed, depending on molecular chirality.

In this work, we present the result of a comprehensive experimental exploration of the nematic states in two different families of dimeric materials. The main result is that in addition to the N phase, both classes feature an $N_{tb}$ phase with the local director (defined as the average orientation of the dimeric arms) shaped as an oblique helicoid, Fig.1b. The oblique helicoidal structure of $N_{tb}$ is evidenced by FFTEM textures of Bouligand arches[23] of two distinct types. The local director is modulated along the helicoidal axis with a period of about (8-9) nm, which is 2-3 orders of magnitude shorter than typically found in the chiral N* phase. The tendency of the molecules to form local bend-twist configurations as a condition of the N-$N_{tb}$ transition is confirmed by the temperature dependence of the bend modulus $K_3$ measured in the N phase: $K_3$

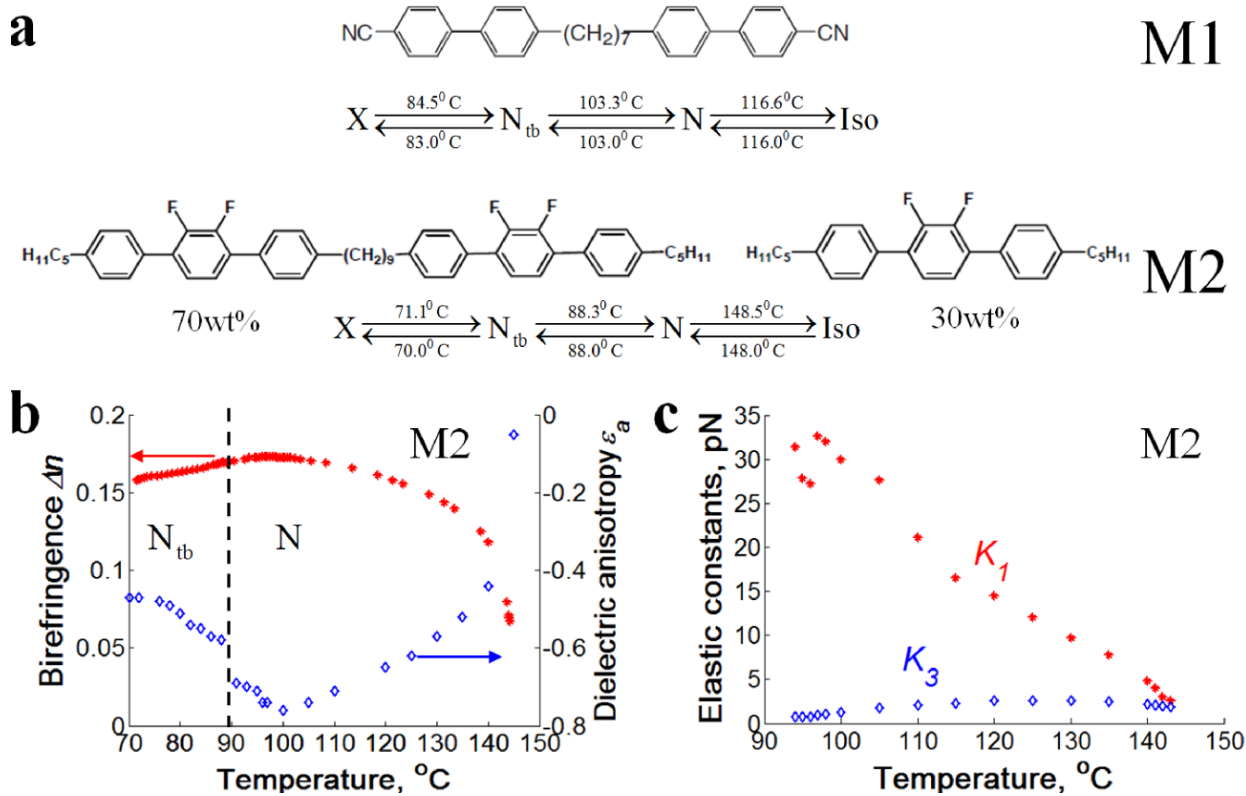

**Figure 2 | Properties of $N_{tb}$ materials.** (a) Structural formulae and phase diagrams of M1 and M2. (b) Temperature dependence of birefringence and dielectric anisotropy of M2. (c) Temperature dependence of $K_1$ and $K_3$ for M2.

decreases near the transition into the $N_{tb}$ phase to the anomalously low levels. Electro-optic response in an alternating current (AC) electric field shows that the field-induced reorientation of $N_{tb}$ allows splay and saddle-splay of the optic axis but not bend nor twist; as a result, the bend Frederiks transition is dramatically different in the two phases. The effect is a natural consequence of the equidistance of layers, $p_{tb} = \text{const}$.

**RESULTS**

**Materials.** The studied materials M1 and M2 are shown in Fig.2a. M1 material, 1",7"-bis(4-cyanobiphenyl-4'-yl)heptane (CB7CB), was synthesized as described in Ref. 24. The dimer has a longitudinal dipole moment at each arm. M1 shows a positive dielectric anisotropy $\varepsilon_a = \varepsilon_\parallel - \varepsilon_\perp > 0$ in the N phase (the subscripts indicate directions parallel to $\hat{\mathbf{n}}$ and perpendicular to it). M2 is a mixture, of a dimer 1,1,1-Di(2',3"-Difluoro-4-pentyl[1,1';4',1"]terphen-1"-yl)undecane (DTC5C9) (for synthesis, see Methods) and a monomer MTC5 (added in order to improve alignment, reduce viscosity and working temperatures); weight proportion is DTC5C9 (70wt%):MTC5(30%). The dielectric anisotropy of M2 is negative, $\varepsilon_a < 0$, which allows us to illustrate a dramatic difference of elastic properties and dielectric response of N and $N_{tb}$ by exploring the bend Frederiks transition[22] between the homeotropic and distorted state of the optic axis in both phases.

**Optical textures and birefringence.** M1 and M2 show a similar phase diagram: A uniaxial N at high temperatures $T$ and a different phase $N_{tb}$ at lower $T$. In cells with homeotropic alignment (achieved by an inorganic passivation layer NHC AT720-A Nissan Chemical Industries, Ltd.), the N texture observed between two crossed polarizers is uniformly dark, as it should be, as the optic axis $\hat{\mathbf{n}}$ is along the direction of observation. In the $N_{tb}$ state, the texture remains dark, i.e., the material remains optically uniaxial.

In planar cells (aligned by rubbed polyimide PI2555 films (HD Microsystem)), the phase transition N-$N_{tb}$ upon the temperature decrease in both M1 and M2 is evidenced by a propagating front after which a texture of stripes is established. These range from faint stripes of sub-µm scale to macroscopic stripes, (10-100) μm wide. These stripes are not thermodynamically stable. By applying an AC electric field of a frequency 10 kHz (parallel to $\hat{n}$ in M1 and perpendicular to $\hat{n}$ in M2), we eliminate the stripes to achieve an optically homogeneous state. If the field is removed and the temperature is fixed or raised, the stripe pattern does not reappear.

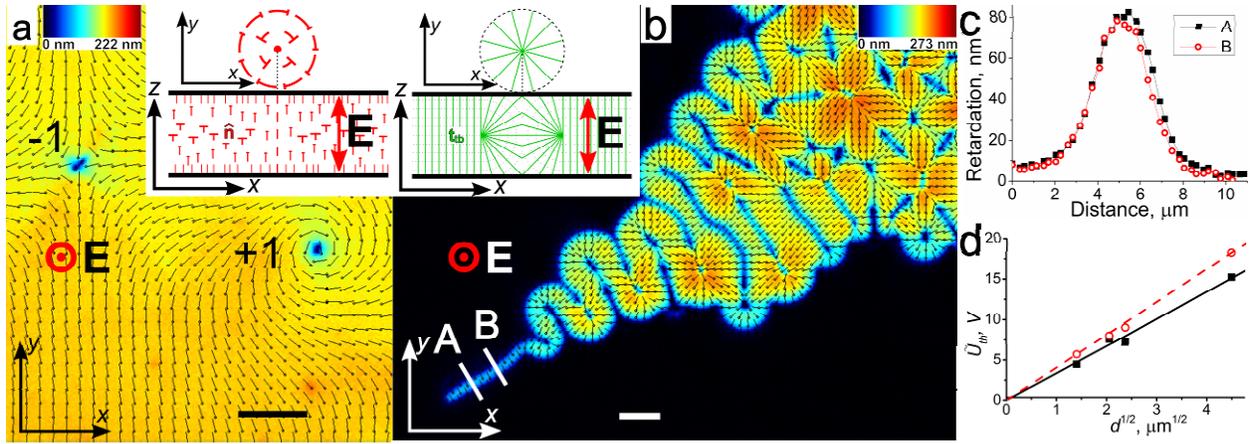

**Figure 3 | Dielectric reorientation of the optic axis in the plane of homeotropic M2 cells. (a)** PolScope texture of bend Frederiks transition in the N phase, showing the local optic axis projected onto the cell's plane (black bars) and the local optical retardation of the cell (pseudocolors). Scale bar 10 μm. Inset shows bend-twist of the director (red symbols; nails depict titled director). **(b)** PolScope texture of bend Frederiks transition in the $N_{tb}$ phase showing only splay and saddle-splay deformations of the optic axis. The inset shows the initial nucleating site in the shape of an axially symmetric toroidal focal conic domain[29] with green lines following the optic axis. Scale bar 10 μm. **(c)** Optical retardation of the realigned $N_{tb}$ structure measured along the lines A and B shown in (b). **(d)** Threshold voltage of expansion of a realigned $N_{tb}$ vs cell thickness $d$, for $T = 87°C$ (filled symbols) and $T = 86°C$ (open symbols).

These uniform states were used to determine birefringence $\Delta n = n_e - n_o > 0$ in both phases, Fig.2b. Equation (1) predicts that $\Delta n$ decreases in the $N_{tb}$ phase, by a factor $\left(1 - 3\theta_0^2 / 2\right)$, see Methods. For M2, the decrease of $\Delta n$ is about 7% from its maximum value in the N phase, which allows us to estimate the tilt as $\theta_0 \approx 17°$ at 79°C. If the temperature of the homogeneous sample is reduced, the stripes typically reappear. This behavior is consistent with a Helfrich-Hurault undulation (buckling) instability observed in smectics and N*, and caused by the temperature-induced decrease in layers periodicity (pitch)[22]. We conclude that the true layered nature of $N_{tb}$ must be associated with (optically) invisible submicron features.

**Elastic constants in N phase.** Uniform planar and homeotropic alignment was used to determine the elastic constants of M2 by the Frederiks effect, i.e., material reorientation by an applied 10 kHz electric field (for the elastic constants of a similar material, see Ref. 25). When the field is parallel to $\hat{\mathbf{n}}$ in the homeotropic N cell, $\hat{\mathbf{n}}$ starts to tilt above some threshold voltage[26] $U_{th} = \pi\sqrt{K_3 / (\varepsilon_0 |\varepsilon_a|)}$, which yields the value of $K_3$, since $\varepsilon_a$ is known, Fig.2b. The splay constant $K_1$ is determined by the threshold fields of realignment in planar cells[27]. Behavior of $K_1$ is typical for N materials, while that of $K_3$ is not, as $K_3$ decreases to a very low value 0.77 pN as the temperature is lowered towards the N-$N_{tb}$ transition, Fig.2c; similarly small $K_3$ was recently measured for a mixture of dimers[20].

**Dielectric reorientation of optic axis in N and $N_{tb}$.** The homeotropic cells of M2 with a vertical AC field allow us to trace an important difference in the Frederiks reorientation of the N and $N_{tb}$ phases. In the N phase, Fig.3a, once $U > U_{th}$, the optic axis realigns gradually and everywhere, as in the second-order transition. Since the tilt direction is degenerate, it results in umbilics, i.e., defects of winding numbers -1 and +1, Ref.28. The +1 umbilics show an in-plane bend of $\hat{\mathbf{n}}$, which is expected, as $K_3 \ll K_1$. The same experiment in $N_{tb}$ reveals a completely different scenario. Reorientation of the optic axis starts only at isolated sites of the sample, associated with dust particles or surface irregularities. The nucleating regions in the shape of axisymmetric focal conic domains[29] coexist with the homeotropic surrounding (inset in Fig.3b); they expand if the voltage is higher than some threshold $\tilde{U}_{th}$. The deformations of the optic axis are of splay and saddle splay type, Fig.3b.

The profile of optical retardation measured across the domain of reoriented $N_{tb}$ reaches a maximum at the center of the domain, Fig.3c, indicating that the tilt of the optic axis is at a maximum in the center. Thus the pattern is similar to the field-induced reorientation in smectic A (Ref.29) and $N^*$ (Ref.30) phases with $\varepsilon_a < 0$, in which the layers equidistance allows splay and saddle-splay, but prohibits bend and twist of the layers' normal. The threshold voltage $\tilde{U}_{th}$ of expansion in layered liquid crystals is determined mostly by the balance of surface anchoring at the plates and the dielectric reorienting torque, so that[29,30] $\tilde{U}_{th} = 2\sqrt{\dfrac{dW}{\varepsilon_0 |\varepsilon_a|}} \propto \sqrt{d}$, where $W$ is the surface anchoring strength and $d$ is the cell thickness; the dependence $\tilde{U}_{th} \propto \sqrt{d}$ agrees with the experiment, Fig.3d. In the N phase, the Frederiks voltage threshold does not depend on $d$. The peculiar character of the dielectric response provides another argument in favor of periodic nature of $N_{tb}$ at the scales much shorter than the visible scales.

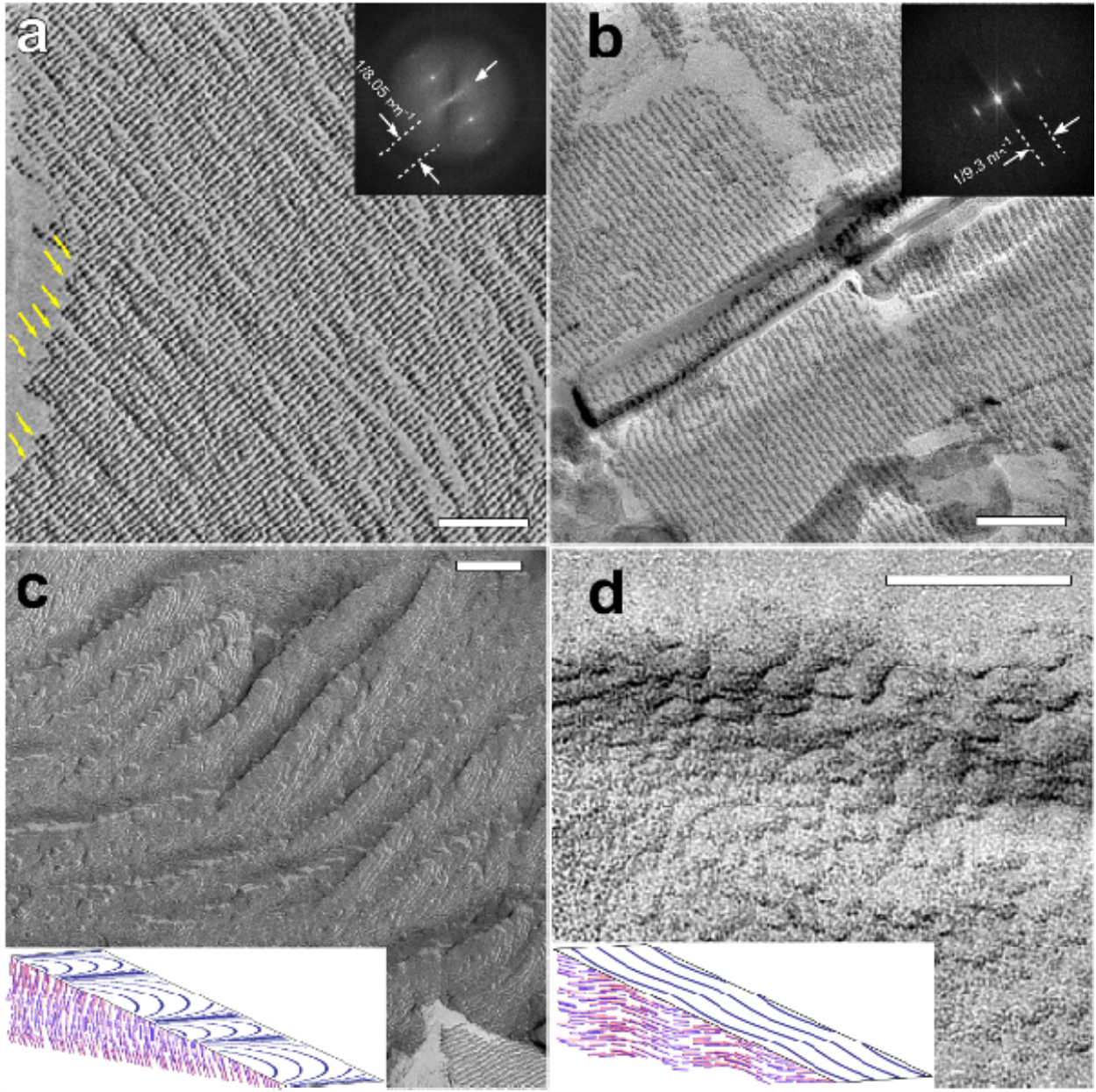

**Figure 4 | FFTEM textures of $N_{tb}$ with uniform and arched structures. (a, b)** FFTEM textures and corresponding fast Fourier transform (FFT) patterns of (a) M1 with pitch $p_{tb} = 8.05$ nm and (b) M2 with $p_{tb} = 9.3$ nm, viewed in the planes parallel to the optic axis. The arrows in (a) point towards domain boundaries of average extension 26 nm, which are roughly perpendicular to the $N_{tb}$ layers. Presence of domains is also revealed by a diffuse intensity pattern in FFT, marked by a white arrow in (a). **(c, d)** FFTEM

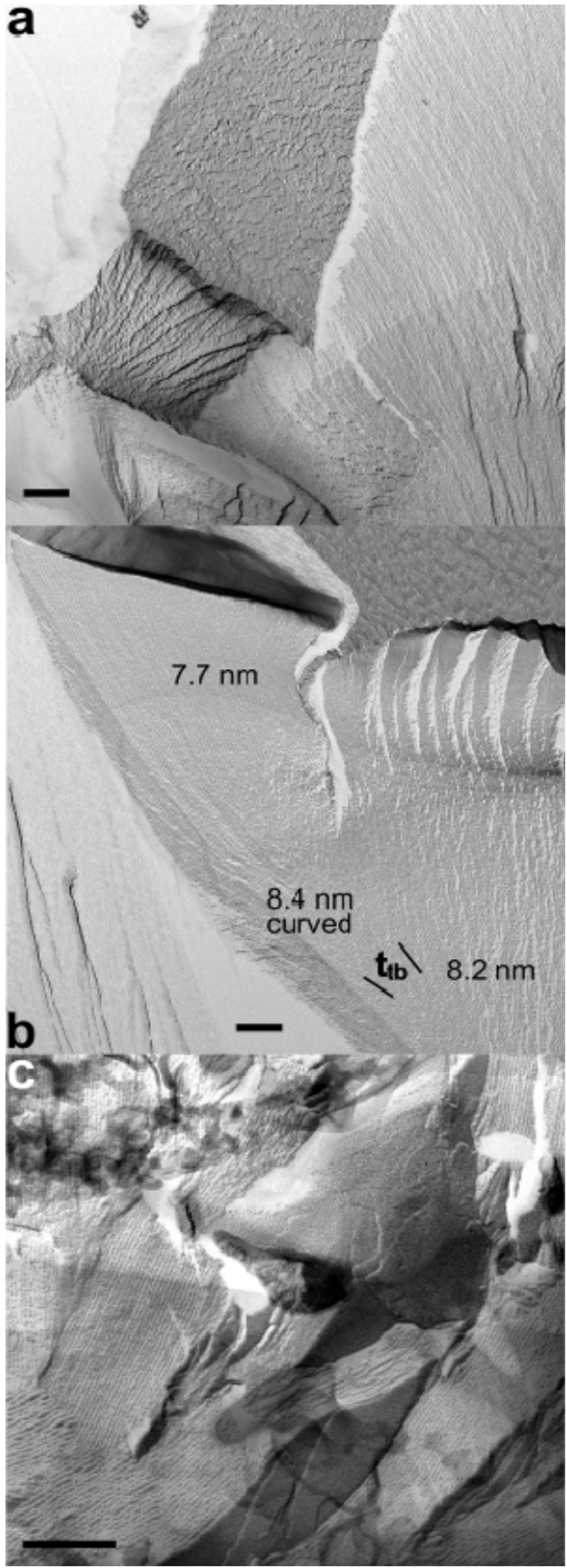

image of Bouligand arches in M1 formed as imprints of the oblique helicoidal structure onto the fracture plane that is (c) almost perpendicular and (d) almost parallel to the helicoid axis of $N_{tb}$. The insets show the corresponding schemes of Bouligand arches of two types in $N_{tb}$, calculated for (c) $\theta_0 = 17°$, $\psi = 5°$ and (d) $\theta_0 = 20°$, $\psi = 57°$. All scale bars are 100 nm.

**Figure 5. | FFTEM textures of non-uniform $N_{tb}$ samples.** **(a)** M1 (CB7CB) quenched at 95°C, showing layered structures of different orientations. The period of most domains varies between 8.0-8.2 nm. Material: M1. **(b)** M1 with splay distortions of the helix axis and atypical periodicity of 7.7 nm. Enlarged portions of the texture are displayed in Supplementary Figures S1 and S2. **(c)** M2, exhibiting domains with layered structure oriented along different directions. The period of the layered structure varies between 8.6-10.3 nm. Scale bar 200 nm in all images.

**Nanoscale periodic arrangement of molecular orientation.** The layered structure of $N_{tb}$ phase is clearly evidenced in FFTEM images of Pt/C replicas of fractured M1 (Fig.4a, Fig.5a,b), and M2 (Fig.4b, Fig. 5c). Most of them show a one-dimensional layered structure with a period ~(8–9) nm, corresponding to the pitch $p_{tb}$ of director deformations, in agreement with the findings by Chen et al for M1.[19] In M1, the regularly observed value of $p_{tb}$ is 8.05 nm, Fig.4a, while in M2, $p_{tb}$ = 9.3 nm, Fig.4b. Frequent observation of layers that are perpendicular to the fracture plane, Fig.4a,b, correlates with the theoretical predictions and experiments[31] on freeze-fractured N and N$^*$, in which the fracture plane tends to be parallel to $\hat{\mathbf{n}}$, as it minimizes the density of molecules in the cut surface. In the $N_{tb}$ phase, the surface with the minimum molecular density is not flat, but modulated with a period $p_{tb}$. This leads to a shadowing effect and explains why oblique deposition of the Pt/C film yields the period $p_{tb}$, Fig.6a. On rare occasions, FFTEM textures of M1 exhibit rather unusual patterns with a period smaller than 8 nm, e.g., 7.7 nm (Fig.5b), 7.4, 4.8, and 3.4 nm (Supplementary Figure S1). It is unlikely that all of these small periods can be explained by absence of shadowing effect when the oblique deposition direction happens to be perpendicular to the wave-vector $\mathbf{t}_{tb}$ of the $N_{tb}$ helix, Fig.6b. Most likely, these small periods are associated with packing of different conformers that are known[32] to exist in dimer materials similar to M1 and M2. The energy difference between the conformers is very small[32], thus some of them might form twist-bend structures with the period different from $p_{tb}$ = (8-9) nm of the prevailing conformer.

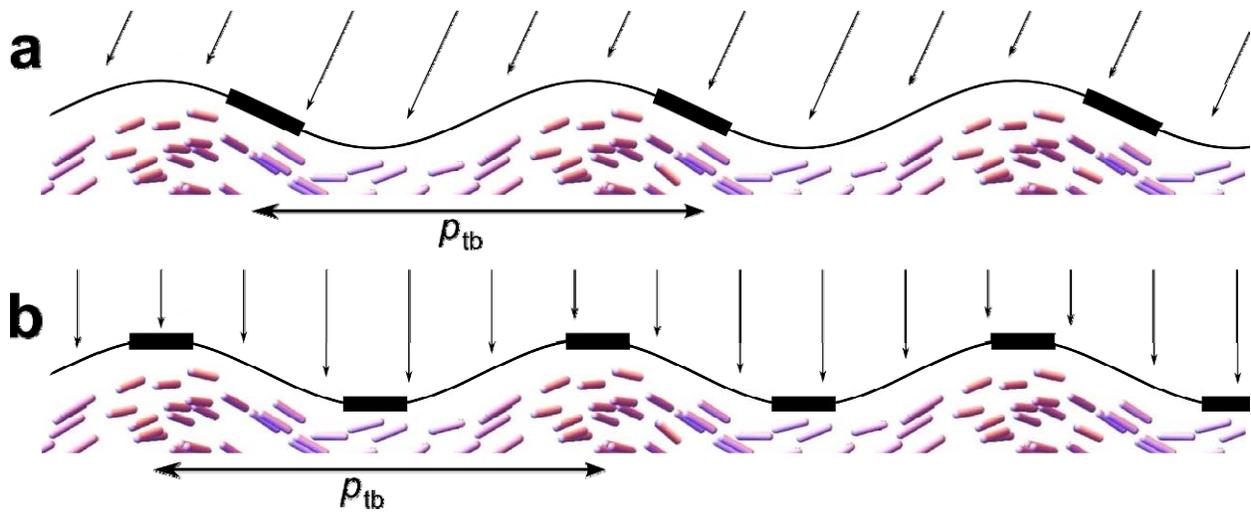

**Figure 6. | Scheme of oblique deposition onto the fracture surface of $N_{tb}$.** (a) General case, the fracture surface corresponds to minimum density of molecules; period of replica $p_{tb}$; (b) Hypothetical limiting case, period of replica is two times smaller than the helix pitch; deposition direction is orthogonal to the wave-vector of the helix.

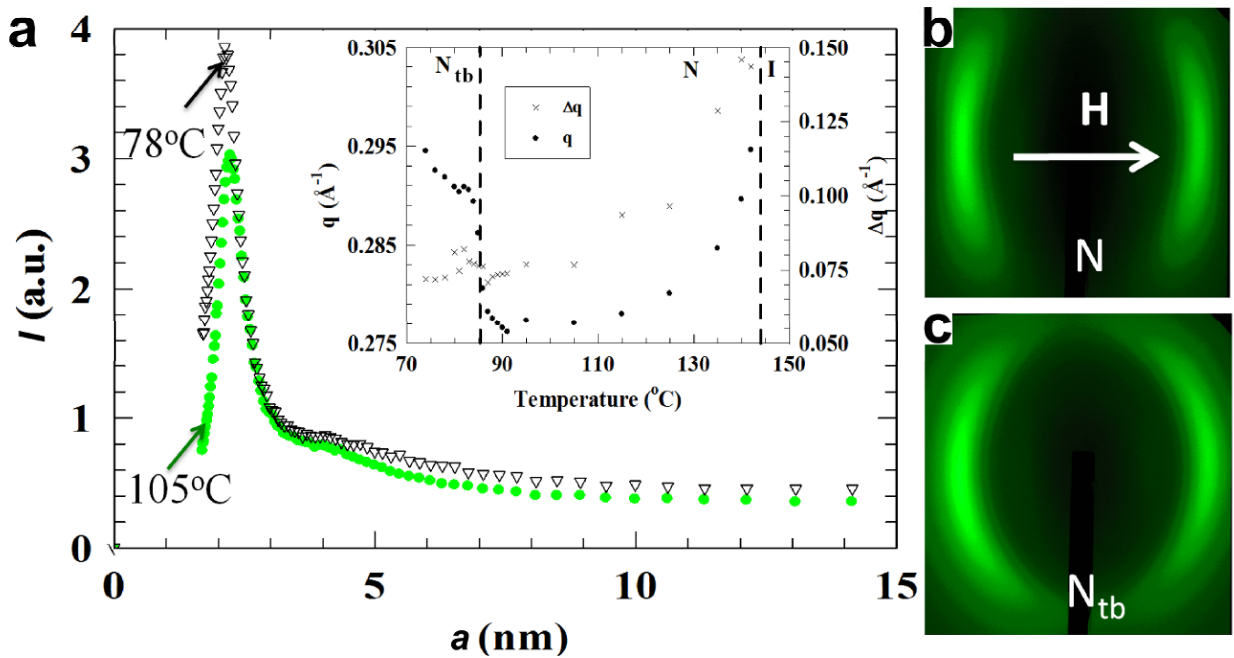

**Figure 7 | XRD results for M2 in the N and $N_{tb}$ phases.** (a) Typical dependence of diffraction intensity vs. wave-vector $q$, presented in terms of the length scale $a$; inset

shows the temperature dependences of $q$ and $\Delta q$, the full width at half maximum of scattered intensity; **(b)** 2D scattering pattern for the N phase; **(c)** 2D scattering pattern for the $N_{tb}$ phase.

Our XRD study shows that the pitch of (8-9) nm of the $N_{tb}$ phase in M2 is not associated with the smectic-like modulations, as the intensity of X-ray scattering is featureless in the range 5 nm-14 nm, Fig.7a. Smaller repeat distances of ~2.2 nm are observed in both N and $N_{tb}$ phases, but the correlation length of these is relatively small, up to 9 nm, Fig.7a, indicating that the long-range structure of both N and $N_{tb}$ is nematic, Fig.7b,c, rather than smectic-like. One should not exclude the possibility of cybotactic clusters, embedded into the $N_{tb}$ phase. Recently, Meyer, Luckhurst and Dozov[18] explored the flexoelectric effect in M1 and suggested that its features are consistent with an oblique helicoidal structure with $p_{tb}$ = 7 nm, if one assumes standard values of the flexoelectric coefficients. This estimate is very close to the periodicities directly seen in Fig.4a,b, 5, and Supplementary Figures S1 and S2.

**Oblique helicoidal geometry of director.** The second important type of FFTEM textures is that one of periodic arches, Fig.4c,d. These arches are very different from the celebrated Bouligand arches[23] of the cholesteric $N^*$ liquid crystals. In $N^*$, each arch corresponds to a rotation of $\hat{n}$ by $\pi$ and any two adjacent arches are indistinguishable from each other, as should be for a right-angle helicoid. In the $N_{tb}$ phase, the geometry is very different, Fig.4c,d and Fig.8, since the underlying structure is an oblique helicoid rather than a right-angle helicoid. Little is known that in the Appendix of the original paper[23], in addition to the $N^*$ arches, Bouligand also considered asymmetric arches for a hypothetical fractured system of oblique helicoids. We extend his approach to the specific case of Equation (1), written for the unit director field. Suppose that the plane of fracture ($xy'$) is tilted around the axis $x$, Fig.1b, by an angle $\psi$ measured between the new axis $y'$ and the original axis $y$. The director components in the fracture plane are $n_x = \cos t_{tb} z \sin \theta_0$ and $n_{y'} = \sin t_{tb} z \sin \theta_0 \cos \psi - \cos \theta_0 \sin \psi$. The local orientation of the director projection in the obliquely fractured N$_{tb}$ phase is then given by the equation $\dfrac{dx}{dy'} = \dfrac{\cos(t_{tb} y' \sin \psi) \sin \theta_0}{\sin(t_{tb} y' \sin \psi) \sin \theta_0 \cos \psi - \cos \theta_0 \sin \psi}$, with the solution

$$x - x_0 = \frac{\ln \left| 2 \cos \psi \sin \theta_0 \left[ \sin(t_{tb} y' \sin \psi) - \tan \psi \cot \theta_0 \right] \right|}{t_{tb} \sin \psi \cos \psi}. \qquad (2)$$

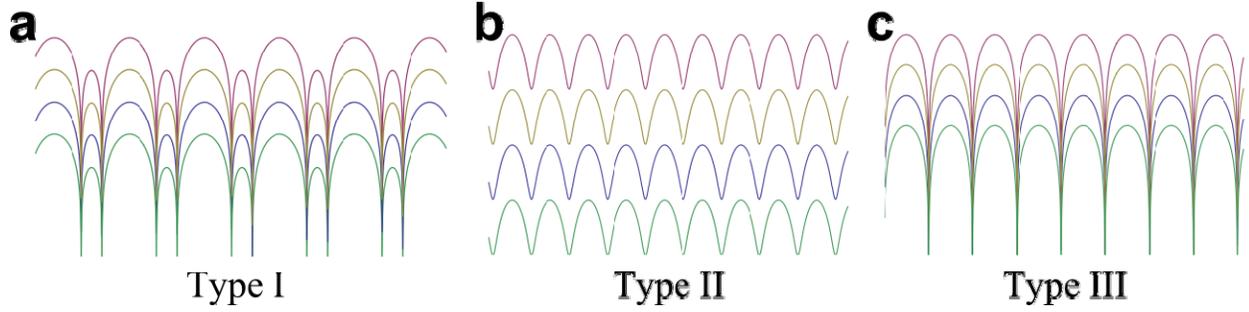

**Figure 8 | Three types of Bouligand arches predicted by Equation (2). (a)** type I, alternating wide and narrow arches, with $\theta_0 = 0.3$, $\psi = 0.1$; **(b)** type II, $\theta_0 = 0.3$, $\psi = 0.35$; **(c)** type III, $\theta_0 = \psi = 0.3$.

Here $x_0$ is the shift of one arch with respect to the other. Equation (2) distinguishes three types of the Bouligand arches that should be observed in a material with oblique helicoidal structure: type I for $\psi < \theta_0$, Fig. 8a, with alternating wide and narrow arches, type II for $\psi > \theta_0$, with a wavy structure, Fig.8b, and intermediate type III, with $\psi = \theta_0$, that is hard to distinguish from the classic symmetric N* arches, Fig.8c. The type I and II arches, never seen in the chiral nematic N* phase, are readily distinguishable in the textures of the N$_{tb}$ phase, Fig.4c and 4d, respectively. In type I, inset in Fig.4c and Fig.8a, the director imprint rotates in the entire range (0-2$\pi$) of azimuthal angles in the fracture plane, but the odd and even arches are of a different width, $l_{0,\pi} \neq l_{\pi,2\pi}$. Type II represents a wavy structure that extends in the direction normal to the helicoidal axis but do not explore the entire range of azimuthal orientations, Fig.4d, Fig.8b and Supplementary Figure S3, in contrast to other types of Bouligand arches where the azimuthal reorientation is by $\pi$ within each arch. Observation of type I and type II arches provides a clear evidence of the oblique helicoidal structure of the $N_{tb}$ phase.

**Nonuniform textures of $N_{tb}$.** FFTEM textures show that the $N_{tb}$ structure is sometimes modulated not only in the direction of helix axis, but also along other directions, in particular, in the *xy* plane perpendicular to the helix axis, Fig.4a, Fig.5, Supplementary Figures S1, S2, and S3. For example, in Fig.4a, the periodic arrangements of twist-bend director have a limited width of about 20-30 nm. The domains are separated by structure-less boundaries, with no clear periodicity on the scales of $p_{tb}$, apparently of an N type. The modulation might be caused by defects such as grain boundaries, screw dislocations parallel to the helicoidal axis, and by coexistence of left-handed and right-handed twisted domains. Homochiral domains have been indeed observed in electro-optical studies[12,17,18], but at a much larger (supramicron) scale. One should expect that the spatial extension of the homochiral domains is determined by the kinetic history of sample preparation, confinement conditions, and other factors, such as presence of the electric field[12]. Further exploration is needed to understand the exact mechanisms behind the modulated structures seen in Fig.4a. Such a work is in progress.

Besides the domain textures of $N_{tb}$ with abruptly changing orientation of the helix axes $\mathbf{t}_{tb}$, Fig.5a, c, one also observes regions with smooth splay type reorientation of $\mathbf{t}_{tb}$, marked in

Fig. 5b. Splay deformation seen in FFTEM textures at the nanoscale is consistent with the idea that the deformed $N_{tb}$ structure tends to preserve the pitch of helicoid and with the observation of splay and saddle-splay deformations on the optical (micrometer) scales, Fig.3b. Predominance of splay and saddle splay in the distorted configurations of $\mathbf{t}_{tb}$, combined with the very small (nanometers) period of the oblique helicoidal structure makes the polarizing microscope textures of the $N_{tb}$ phase very similar to those of the smectic phases.

**DISCUSSION**

The structural, elastic, optical, dielectric and electrooptical properties of the two different families of dimer compounds clearly demonstrate the existence of the $N_{tb}$ phase with a local twist-bend structure. The results underscore a complex interplay between the flexible nature of the achiral dimeric molecules with aliphatic chains containing an odd number of methylene groups, and their chiral nanoscale organization. The prevailing element of the twist-bend nematic order is an oblique helicoid, Fig.1b, formed by the local director, associated with the average orientation of the dimers' arms. The obliquie helicoid structure reveals itself in the unique shape of the Bouligand arches that are either asymmetric (type I, Fig.4c, Fig.8a) or not fully developed in the sense of director rotations in the fracture plane (type II, Fig.4d, Fig.8b and Supplementary Fig.S3). These two types of arches are different from the classic cholecteric arches that are always symmetric and fully developed[23]. The type I and II arches were originally proposed by Bouligand as a hypothetical imprint of chiral fibrilles in chromosomes of Dinoflagellates[23], but were not observed so far.

The TEM measurements reveal that the periodic modulations of the director along the helical axis has a very short period, about 8-9 nm for both studied materials M1 and M2. These data compare well to the TEM data by Chen et al[19] and estimates based on electro-optic response[18] of M1. Although one does observe a typical period of (8-9) nm, there are also examples of much shorter periodicity, from 7.7 nm to 3.4 nm. Another aspect of nanostructural organization that deserves further studies is a modulation in the direction more or less perpendicular to the helicoidal axes, that can be caused by structural defects such as screw dislocations and by coexisting left- and right-twisted domains.

By exploring the dielectric response, we demonstrated that the classic Frederiks effect in the homeotropic cells is very different when staged in the normal N phase and in the $N_{tb}$ phase, because of the tendency of twist-bend director modulations to keep equidistance. The temperature dependence of the bend modulus and its very low value near the N-$N_{tb}$ phase transition put a new challenge to our understanding of molecular mechanisms of elastic properties of liquid crystals. A closely related issue is the relationship between the twist-bend $N_{tb}$ structure and double-twist structure of the so-called blue phases that are known to be stabilized by the dimeric molecules[33]. Finally, practically nothing is known about the hydrodynamics of the $N_{tb}$ phase (apart from the fact that it is much more viscous than its high-temperature N neighbour). Further studies of the $N_{tb}$ phase promise a dramatic improvement of our understanding of the long-range orientational order which shows new intriguing facets at the nanoscale.

*Note added in proofs.* *After this paper hasd been submitted and reviewed, we learned that a modulated orientation of molecules with a period of 14 nm, consistent with the structure of the $N_{tb}$ phase, has been observed in the bent-core nematic by Chen et al [34].*

# METHODS

**Synthesis of DTC5C9 1,5-Bis(2',3'-difluoro-4''-pentyl-[1,1':4',1''-terphenyl]-4-yl)nonane**

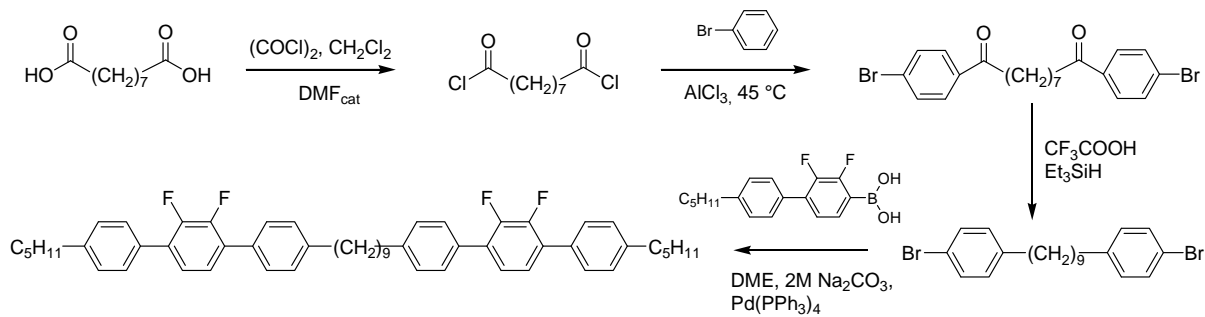

Preparation of nonanedioyl dichloride (1): Nonanedioic acid (17 g, 0.09 mol) was dissolved in dry dichloromethane (200 ml) under nitrogen. Oxalyl chloride (30 ml, 44.3 g, 0.35 mol) was added followed by dry *N, N'*-dimethylformamide (1 drop) and gentle stirring. The nitrogen supply was turned off and the reaction was monitored by the evolution of gas. The reaction was stirred overnight and then refluxed for 2 hours until gas evolution had ceased. The solution was evaporated by vacuum distillation to yield crude nonanedioyl dichloride which was used immediately in the Friedel-Crafts acylation described below. Due to the instability of acyl chlorides the crude product was used immediately; no NMR spectra were recorded.

Preparation of 1,7-bis(4-bromophenyl)nonane-1,7-dione (2): Bromobenzene (50 ml, 74.75 g, 0.47 mol) was added to powdered aluminium chloride (25 g, 0.187 mol) and rapidly stirred under nitrogen for 1 hour. The crude nonanedioyl dichloride was dissolved in bromobenzene (30 ml, 44.85 g, 0.285 mol) and was added dropwise over a period of one hour. The temperature of the reaction was maintained below 45 °C during the addition. The reaction mixture was maintained at 40 °C and stirred overnight. The reaction was allowed to cool to ambient temperature and poured into a solution of concentrated hydrochloric acid (150ml) in ice water (600 ml). A small portion of dichloromethane was added to dissolve the suspended solids. The aqueous layer was extracted with dichloromethane several times and the organic extracts were dried over magnesium sulphate with rapid stirring overnight. The dichloromethane was evaporated under reduced pressure. The crude solid was recrystalised from ethanol four times to yield 1,5-bis-(4-bromophenyl)nonane-1,5-dione. The yield was found to be 18.4 g, 44 %

**¹H NMR (CDCl₃**, *J*/Hz, 400 MHz): δ 7.81 (4H, d, *J* = 8.6); 7.59 (4H, d, *J* = 8.6); 2.92 (4H, t, *J* = 7.3); 1.69-1.76 (4H, m); 1.42-1.35 (6H, m).

Preparation of 1,5-bis-(4-bromophenyl)nonane (3): 1,5-Bis-(4-bromophenyl)heptane-1,5-dione (8.8 g, 0.02 mol) was dissolved in trifluoroacetic acid (30.75 ml, 45.79 g, 0.40 mol). Triethylsilane (16.6 ml, 12.09 g, 0.10 mol) was added dropwise over a period of one hour with rapid stirring and cooling in a water bath. The reaction mixture became turbid white after 1 hour. The reaction was stirred at ambient temperature for 48 hours. The reaction mixture was poured into a mixture of ice and water. Hexane was added and the aqueous layer was extracted in hexane four times. The organic extracts were combined and were dried over magnesium sulphate overnight. The solvent was then evaporated. The crude product was separated by column (silica) chromatography and yielded 1,5-bis-(4-bromophenyl)heptane. The mobile phase was a hexane/dichloromethane gradient starting with pure hexane finishing at 35 % dichloromethane. The yield was found to be 3.17 g, 36.2 %

¹H NMR (CDCl₃, *J*/Hz, 400 MHz): δ 7.38 (4H, d, *J* = 8.4); 7.04 (4H, d, *J* = 8.6); 2.55 (4H, t, *J* = 7.7); 1.53-1.61 (4H, m); 1.24-1.35 (10H, m).

Preparation of 1,5-bis(2',3'-difluoro-4''-pentyl-[1,1':4',1''-terphenyl]-4-yl)nonane (4): A solution of 1,5-bis-(4-bromophenyl)nonane (0.48 g, 0.01 mol) and (2,3-difluoro-4'-pentyl-[1,1'-biphenyl]-4-yl)boronic acid (1.0 g, 0.03 mol) in a mixture of 1,2-dimethoxyethane (12 ml) and saturated aqueous solution of sodium hydrogen carbonate (18 ml) was thoroughly degassed with argon for 1 hour. Tetrakis(triphenylphosphine)palladium(0) 50 mg was added. The reaction was refluxed at 125 °C for 12 hours under stirring. The reaction was allowed to cool to room temperature. Water was added to the reaction. The aqueous solution was extracted with dichloromethane four times and dried over magnesium sulphate overnight. The solvent was evaporated and the crude product was separated by column chromatography and yielded 1,5-bis(2',3'-difluoro-4''-pentyl-[1,1':4',1''-terphenyl]-4-yl)nonane. The mobile phase was a hexane/dichloromethane gradient starting with pure hexane, finishing at 50 % dichloromethane. The yield was found to be 0.66 g, 70 %.

$^1$H NMR (CDCl$_3$, $J$/Hz, 400 MHz): δ 7.50 (8H, d, $J$ = 8.0); 7.28 (4H, d, $J$ = 8.0); 7.22 (4H, m); 2.64 (8H, t, $J$ = 8.0); 1.62-1.70 (8H, m); 1.39-1.38 (18H, m); 0.91 (6H, t, $J$ = 7.1).

The compound MCT5 (CAS registry number 121218-76-6) is available from Kingston Chemicals Ltd, Hull, UK.

**Sample preparation for optical and electro-optical studies.** Experimental cells were assembled from parallel glass plates with transparent indium tin oxide (ITO) electrodes. For planar (tangential) alignment, the substrates were spin coated with polyimide PI2555 HD MicroSystems; homeotropic alignment was achieved an inorganic passivation layer NHC AT720-A Nissan Chemical Industries, Ltd. The temperature was controlled with the Linkam controller TMS94 and a hot stage LTS350 (Linkam Scientific Instruments) with precision 0.01°C. In all the experiments on dielectric reorientation (Frederiks transitions) of $\hat{\mathbf{n}}$, we used the AC electric field of frequency 10kHz. An AC voltage was applied using function generators DS345 Stanford Research System or Keithley 3390. The voltage was amplified by a wide band amplifier 7602 Krohn-Hite and measured with a Keithley 2000 multimeter. A polarizing microscope OptiPhot2-Pol, Nikon Instruments, Inc. was used for orthoscopic observations. The maps of the in-plane optical axis orientation and phase retardation were obtained by using the Abrio PolScope setup assembled on the basis of Nikon Eclipse E600 Pol microscope.

**Elasticity measurement in the N phase.** The cell thickness used in the experiment was 20 μm for homeotropic cell and 19.9 μm for the planar cell. The cells were filled with the liquid crystal material in the isotropic state and a well aligned N phase was obtained on cooling. For the dielectric characterization, we used an LCR meter HP4284A that measures the capacitance of the sample. The effective dielectric permittivity across the homeotropic cell of thickness $d$ was calculated as $\varepsilon = C \cdot d / (\varepsilon_0 \cdot A)$, where $A$ is the electrode area, and $C$ is the measured capacitance of the cell filled with liquid crystal mixture. The dielectric permittivity parallel to the director, $\varepsilon_\parallel$, was determined from the capacitance measured at low voltages from the homeotropic cell. The perpendicular component $\varepsilon_\perp$ was measured from the planar cell.

To determine the bend elastic constant $K_3$, we followed the Saupe technique, in which one uses a single homeotropic cell and determines the voltage dependence of its capacitance when the material has a negative dielectric anisotropy. The bend elastic constant is obtained by measuring the Frederiks threshold, $K_3 = \varepsilon_0 \cdot |\varepsilon_a| \cdot (U_{th}/\pi)^2$. The Frederiks transition of homeotropic sample was also triggered by the magnetic field directed perpendicularly to the director. The threshold $B_{th3}$ was determined by measuring cell capacitance vs. field strength. Using the expression $K_3 = (d \cdot B_{th3}/\pi)^2 \Delta\chi/\mu_0$, and comparing the bend constant to the value obtained in the electric Frederiks effect, we determined the diamagnetic anisotropy $\Delta\chi = \varepsilon_0 \cdot \mu_0 \cdot |\varepsilon_a| \cdot (U_{th}/(d \cdot B_{th3}))^2$. For example, $\Delta\chi = 2\times 10^{-6}$ at $130°C$.

The elastic constant of splay $K_1$ was obtained by exploring the Frederiks transition of planar sample in the magnetic field. The magnetic field was set normally to the cell and to the planar director; the director reorientation threshold was monitored by measuring the capacitance. The splay elastic constant is defined as $K_1 = (d \cdot B_{th1}/\pi)^2 \Delta\chi/\mu_0$.

**Birefringence.** By averaging the tilted director field in Equation (1) for $\theta_0 > 0$, the effective birefringence of the conical helix in N$_{tb}$ phase is related to the corresponding quantity $\Delta\bar{n} = \bar{n}_e - \bar{n}_o$ in the N phase with unwound helix $\theta_0 = 0$ (presumed to be at the same temperature)

$$\Delta n = \sqrt{\bar{n}_e^2 - (\bar{n}_e^2 - \bar{n}_o^2)\sin^2\theta_0} - \sqrt{\bar{n}_o^2 + \frac{1}{2}(\bar{n}_e^2 - \bar{n}_o^2)\sin^2\theta_0} \approx \Delta\bar{n}\left(1 - \frac{3}{2}\theta_0^2\right).$$

**Sample preparation for freeze fracture-TEM.** To prepare the replica specimens for FFTEM, we put ~0.6 μl of material between two copper planchettes. The sandwich structure was heated (125°C for M1, and 155°C for M2, respectively) to obtain isotropic phase, and then cooled down and kept at deep N$_{tb}$ phase temperature for 5 minutes, 95°C for M1 ($T_{NN_{tb}}^{M1} = 103°C$) and 80°C for M2 ($T_{NN_{tb}}^{M1} = 88°C$). The sample was quenched by plunge freezing in liquid nitrogen, with a high cooling rate >1000°C/s, in order to avoid further phase transitions, and quickly transferred into a freeze-fracture vacuum chamber (BalTec BAF060) where the assembly is kept at -140°C. Inside the chamber, a built-in microtome was used to break the assembly and expose the fractured surface. ~4 nm thick Pt/C was then deposited onto the fractured surface at a 45° angle to create shadowing of the surface structure, followed by a ~20 nm thick C deposition from the top to form a continuous supporting film. The samples were then warmed up and removed from the freeze fracture machine. The liquid crystal material was dissolved in chloroform, while the replica film (often flakes) was picked up and placed onto carbon coated TEM grid and observed using room temperature TEM (FEI Tecnai F20).

**Synchrotron X-ray diffraction studies.** The material was filled into 1 mm diameter quartz tubes located inside a hot stage (Instec model HCS402). The cylindrical neodymium iron boron magnets were used to align the material in the magnetic field 1.5T perpendicular to the incident x-ray beam. Small-angle X-ray scattering was recorded on a Princeton Instruments 2084 × 2084 pixel array CCD detector in the X6B beamline at the National Synchrotron Light Source (NSLS). The beamline was configured for a collimated beam (0.2x 0.3 mm) at energy 16 keV (0.775 Å). In the N phase, there are two diffused peaks centered along the magnetic field, Fig.7b,c, with the wavenumber at the maximum intensity decreasing from $q_o$=2.95nm$^{-1}$ near the clearing point to $q_o$=2.77nm$^{-1}$ at 110ºC (inset in Fig.7a), corresponding to periodicities from $a$=2.16 nm to $a$=2.23 nm. At lower temperatures, a secondary peak at a doubled periodicity 4.46 nm is observed. These two length-scales might correspond to (i) the length of one arm of the dimers and the length of the monomer and (ii) to the length of the entire dimer, respectively. The full width at half maxima ($\Delta q$) is decreasing from $\Delta q = 1.5$ nm$^{-1}$ at 140ºC to $\Delta q = 0.8$ nm$^{-1}$ at 110ºC. This means that the correlation length $\xi = 2\pi/\Delta q$ is increasing from $\xi \approx 4$ nm to $\xi \approx 8$ nm. Such a behavior is typical for an N phase with "cybotactic" smectic clusters, i.e., nanosized clusters of layers with correlation length $\xi$ that increases upon cooling. However, the macroscopic structure is still that one of a fluid N phase. Interestingly, the vertical lobes are nearly straight, indicating no rigid restriction on the layer spacing inside the clusters.

In the N$_{tb}$ phase, $q_o$ increases rapidly from $q_o$=2.77nm$^{-1}$ to $q$=2.93 nm$^{-1}$ (Fig. 7a), which corresponds to local periodicity decreasing from $d$=2.23 nm $to$ $d$=2.14 nm. This can be explained by a bend of dimers with arms tilted away from the straight configuration by about ~15-20º or by the increased mosaicity of the N$_{tb}$ phase. The width of the peak at half maxima is fairly temperature-independent, $\Delta q = 0.75$ nm$^{-1}$, which corresponds to $\xi = 8.4$ nm. It is interesting to see that the lobes of the diffused peaks are much closer to the circular shape, showing that the tilt of the arms of the dimers are much more defined than in the N phase. Such a macroscopically fluid N phase with almost constant smectic nano-clusters is typical of bent-core nematic (BCN) materials.[6]

**REFERENCES**

1   Meyer, R. B. in *Molecular Fluids. Les Houches Lectures, 1973* (eds Balian, R. & Weill, G.)  271-343 (Gordon and Breach, London, 1976).

2   Dozov, I. On the spontaneous symmetry breaking in the mesophases of achiral banana-shaped molecules. *Europhys. Lett.* **56**, 247-253 (2001).

3   Shamid, S. M., Dhakal, S. & Selinger, J. V. Statistical mechanics of bend flexoelectricity and the twist-bend phase in bent-core liquid crystals. *Phys. Rev. E* **87**, 052503 (2013).

4   Memmer, R. Liquid crystal phases of achiral banana-shaped molecules: a computer simulation study. *Liq. Cryst.* **29**, 483-496 (2002).

5   Kamien, R. D. Liquids with Chiral Bond Order. *J. Phys. II France* **6**, 461-475 (1996).

6   Jákli, A. Liquid crystals of the twenty-first century – nematic phase of bent-core molecules. *Liq. Cryst. Rev.* **1**, 65-82 (2013).

7   Ungar, G., Percec, V. & Zuber, M. Liquid crystalline polyethers based on conformational isomerism. 20. Nematic-nematic transition in polyethers and copolyethers based on 1-(4-hydroxyphenyl)2-(2-R-4-hydroxyphenyl)ethane with R = fluoro, chloro and methyl and flexible spacers containing an odd number of methylene units. *Macromolecules* **25**, 75-80 (1992).

8   Silvestri, R. L. & Koenig, J. L. Spectroscopic characterization of trans-gauche isomerization in liquid crystal polymers with two nematic states. *Polymer* **35**, 2528-2537 (1994).

9	Sepelj, M. *et al.* Intercalated liquid-crystalline phases formed by symmetric dimers with an $\alpha,\omega$-diiminoalkylene spacer. *J. Mater. Chem.* **17**, 1154-1165 (2007).

10	Henderson, P. A. & Imrie, C. T. Methylene-linked liquid crystal dimers and the twist-bend nematic phase. *Liq. Cryst.* **38**, 1407-1414 (2011).

11	Panov, V. P. *et al.* Spontaneous Periodic Deformations in Nonchiral Planar-Aligned Bimesogens with a Nematic-Nematic Transition and a Negative Elastic Constant. *Phys. Rev. Lett.* **105**, 167801 (2010).

12	Panov, V. P. *et al.* Field-induced periodic chiral pattern in the $N_x$ phase of achiral bimesogens. *Appl. Phys. Lett.* **101**, 234106 (2012).

13	Cestari, M., Frezza, E., Ferrarini, A. & Luckhurst, G. R. Crucial role of molecular curvature for the bend elastic and flexoelectric properties of liquid crystals: mesogenic dimers as a case study. *J. Mater. Chem.* **21**, 12303-12308 (2011).

14	Cestari, M. *et al.* Phase behavior and properties of the liquid-crystal dimer 1″,7″-bis(4-cyanobiphenyl-4′-yl) heptane: A twist-bend nematic liquid crystal. *Phys. Rev. E* **84**, 031704 (2011).

15	Beguin, L. *et al.* The Chirality of a Twist–Bend Nematic Phase Identified by NMR Spectroscopy. *J. Phys. Chem. B* **116**, 7940-7951 (2012).

16	Imrie, C. T. & Henderson, P. A. Liquid crystal dimers and higher oligomers: between monomers and polymers. *Chem. Soc. Rev.* **36**, 2096-2124 (2007).

17	Panov, V. P. *et al.* Microsecond linear optical response in the unusual nematic phase of achiral bimesogens. *Appl. Phys. Lett.* **99**, 261903 (2011).

18. Meyer, C., Luckhurst, G.R. & Dozov, I. Flexoelectrically Driven Electroclinic Effect in the Twist-Bend Nematic Phase of Achiral Molecules with Bent Shapes. *Phys. Rev. Lett.* **111**, 067801 (2013)

19  Chen, D. *et al.* A Twist-Bend Chiral Helix of 8nm Pitch in a Nematic Liquid Crystal of Achiral Molecular Dimers. Preprint at http://arXiv.org/abs/1306.5504 (2013).

20. Adlem, K., Čopič, M., Luckhurst, G.R., Mertelj, A., Parri, O., Richardson, R.M., Snow, B.D., Timimi, B.A., Tuffin, R.P. & Wilkes, D. Chemically induced twist-bend nematic liquid crystals, liquid crystal dimers, and negative elastic constants. *Phys. Rev. E* **88**, 022503 (2013).

21  Friedel, M. G. Les états mésomorphes de la matière (The mesomorphic states of matter). *Ann. Phys. (Paris)* **18**, 273-474 (1922).

22. Kleman, M. & Lavrentovich, O.D. *Soft Matter Physics: An Introduction.* (Springer, New York, 2003).

23  Bouligand, Y., Soyer, M.-O. & Puiseux-Dao, S. La structure fibrillaire et l'orientation des chromosomes chez les Dinoflagellés. *Chromosoma* **24**, 251-287 (1968).

24  Barnes, P. J., Douglass, A. G., Heeks, S. K. & Luckhurst, G. R. An enhanced odd-even effect of liquid crystal dimers Orientational order in the α,ω-bis(4′-cyanobiphenyl-4-yl)alkanes. *Liq. Cryst.* **13**, 603-613 (1993).

25  Balachandran, R. *et al.* Elastic properties of bimesogenic liquid crystals. *Liq. Cryst.* **40**, 681-688 (2013).

26  Deuling, H. J. Deformation of Nematic Liquid Crystals in an Electric Field. *Mol. Cryst. Liq. Cryst.* **19**, 123-131 (1972).

27    Zhou, S. *et al.* Elasticity of Lyotropic Chromonic Liquid Crystals Probed by Director Reorientation in a Magnetic Field. *Phys. Rev. Lett.* **109**, 037801 (2012).

28    de Gennes, P. G. & Prost, J. *The Physics of Liquid Crystals*. 2nd ed., (Oxford University Press, New York, 1995).

29    Li, Z. & Lavrentovich, O. D. Surface Anchoring and Growth Pattern of the Field-Driven First-Order Transition in a Smectic-A Liquid Crystal. *Phys. Rev. Lett.* **73**, 280-283 (1994).

30    Lavrentovich, O. D. & Yang, D. K. Cholesteric cellular patterns with electric-field-controlled line tension. *Phys. Rev. E* **57**, R6269-R6272 (1998).

31    Berreman, D. W., Meiboom, S., Zasadzinski, J. A. & Sammon, M. J. Theory and Simulation of Freeze-Fracture in Cholesteric Liquid Crystals. *Phys. Rev. Lett.* **57**, 1737-1740 (1986).

32    Hori, K., Iimuro, M., Nakao, A., Toriumi, H. Conformational diversity of symmetric dimer mesogens, $\alpha,\omega$ -bis(4,4'-cyanobiphenyl)octane, -nonane, $\alpha,\omega$ -bis(4-cyanobiphenyl-4'-yloxycarbonyl)propane, and –hexane in crystal structures. *J. Mol. Structure* **699**, 23-29 (2004).

33    Coles, H. J., Pivnenko, M. N. Liquid crystal "blue phases" with a wide temperature range, *Nature* **436**, 99701000 (2005).

34.   Chen, D., Nakata, M., Shao, R., Tuchband, M.R., Shuai, M., Baumeister, U., Weissflog, W., Walba, D.M., Glaser, M.A., Maclennan, J.E., Clark, N.A. Twist-bend heliconical chiral nematic liquid crystal phase of an achiral rigid bent-core mesogen. Preprint at http://arXiv.org/abs/1308.3526 (2013).

**Supplementary Information** is linked to the online version of the paper.

**Acknowledgements**. We thank S. Sprunt, N. Diorio and J. Angelo for the help with XRD. ODL ackowledges useful discussions with N. A. Clark, M. Čopič, I. Dozov, R. D. Kamien, M. Kleman, R. B. Meyer, J. V. Selinger, S. V. Shiyanovskii, and I. Smalyukh. Hospitality of the Isaac Newton Institute for Mathematical Sciences, Cambridge, UK, where part of this work was written, is greatly appreciated. CTI gratefully acknowledges G. R. Luckhurst for the sample M1 and for invaluable discussions. The work was supported by DOE grant DE-FG02-06ER 46331 (electrooptics), NSF DMR grants 1104850 and 1121288, EU FP7 BIND Project 216025. The TEM data were obtained at the (cryo) TEM facility at the Liquid Crystal Institute, Kent State University, supported by the Ohio Research Scholars Program *Research Cluster on Surfaces in Advanced Materials*. Work in Dublin was partly supported by Science Foundation Ireland.

**Author information.** The authors declare no competing financial interests. Correspondence and request for materials should be addressed to ODL (olavrent@kent.edu)

# Nematic twist-bend phase with nanoscale modulation of molecular orientation

V. Borshch, Y.-K. Kim, J. Xiang, M. Gao, A. Jákli, V. P. Panov, J. K. Vij,

C. Imrie, M. G. Tamba, G. H. Mehl, O.D. Lavrentovich

## Supplementary Information

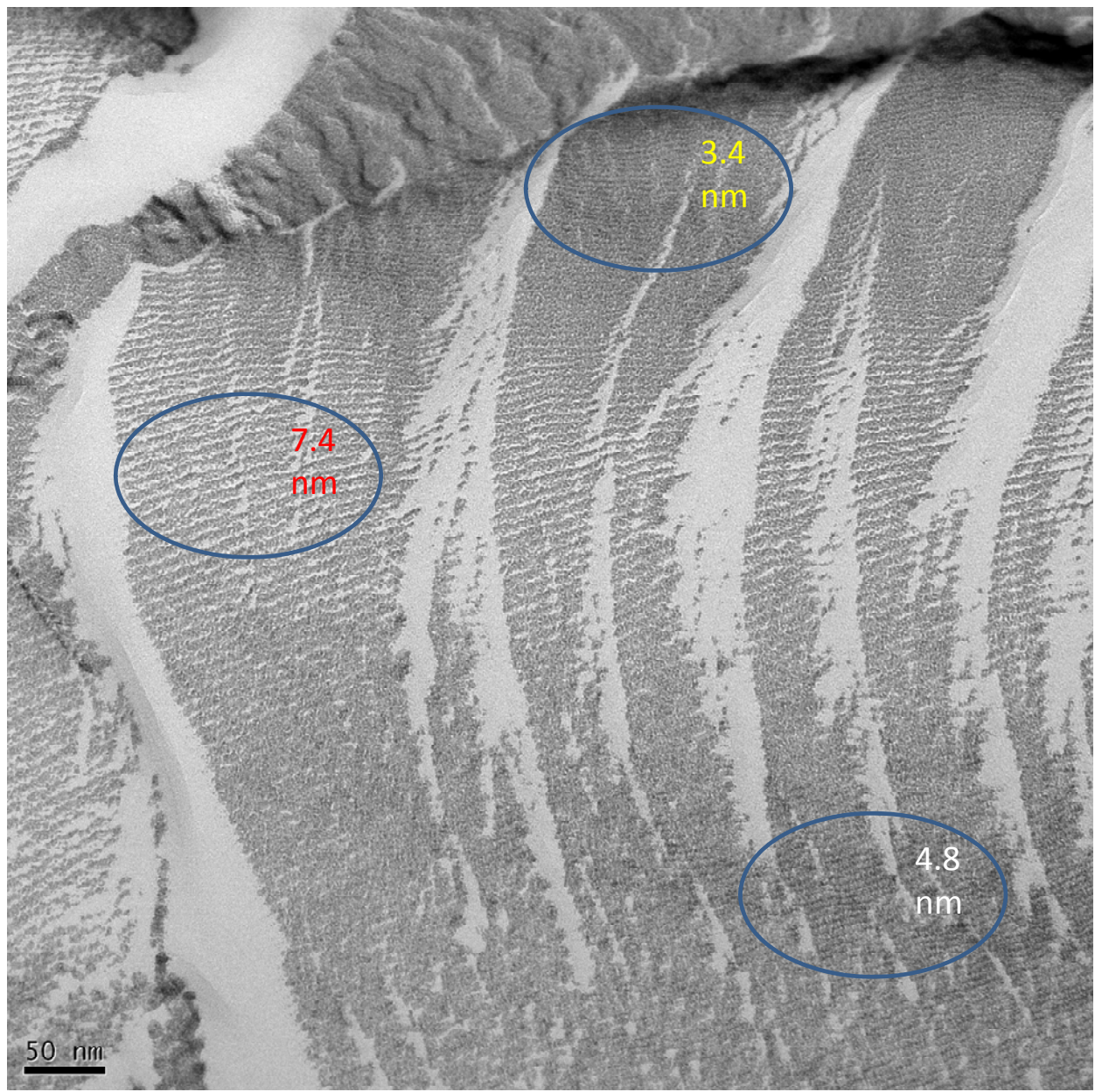

**Supplementary Fig.S1. | FFTEM texture of M1.** Magnified image of a local area in Fig.5b, with unusually small periods of 7.4, 4.8, and 3.4 nm. Scale bar 50 nm.

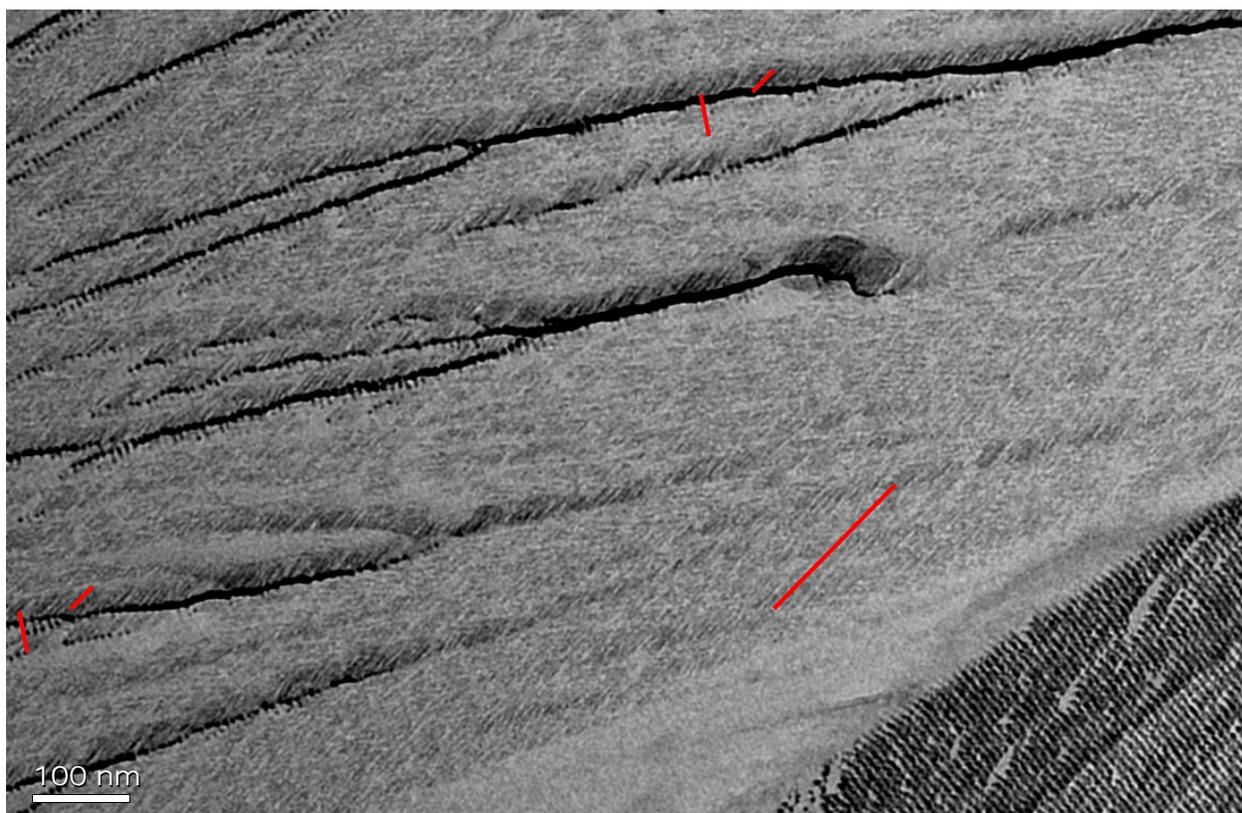

**Supplementary Fig.S2. | FFTEM texture of M1.** Magnified image of a local area in Fig.5b, showing a chiral structure similar to the one observed by Chen et al in Ref.19. Scale bar 100 nm.

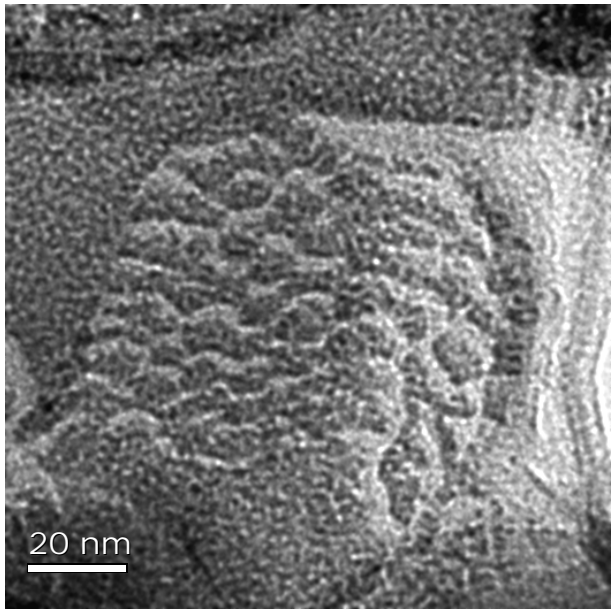 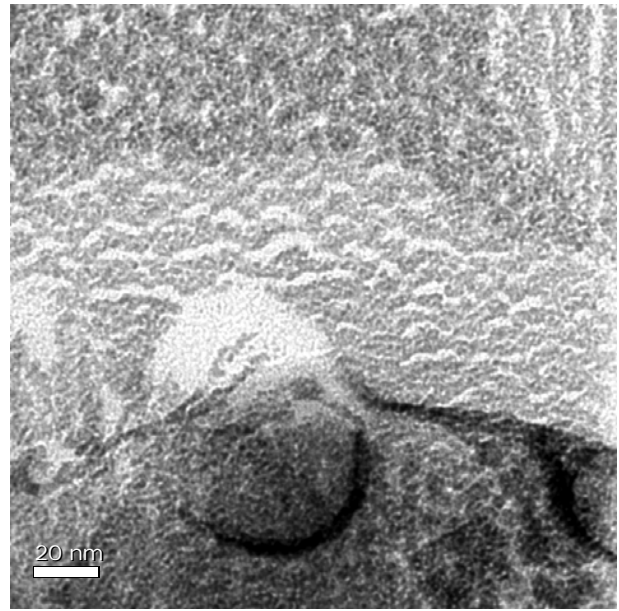

**Supplementary Fig.S3**. | **FFTEM texture of M1.** Enlarged views of Bouligand arches of type II. Scale bars 20 nm.